\documentclass{aa}
\pdfoutput=1
\usepackage[varg]{txfonts}
\usepackage{natbib}
\bibpunct{(}{)}{;}{a}{}{,}
\usepackage{hyperref}
\hypersetup{colorlinks=true,citecolor=blue,linkcolor=blue,urlcolor=blue}
\usepackage{graphicx}
\usepackage{url}
\usepackage{amsmath}
\usepackage{amssymb}
\usepackage{bm}
\usepackage{amsfonts}

\title{Cosmological constraints with weak lensing peak counts and second-order statistics in a large-field survey}

\author{Austin Peel\inst{\ref{inst1}}\thanks{austin.peel@cea.fr}
   \and Chieh-An Lin\inst{\ref{inst1}} 
   \and Fran\c{c}ois Lanusse\inst{\ref{inst2}}
   \and Adrienne Leonard\inst{\ref{inst3}}
   \and Jean-Luc Starck\inst{\ref{inst1}}
   \and Martin Kilbinger\inst{\ref{inst1},\ref{inst4}}}
\institute{Laboratoire AIM, UMR CEA-CNRS-Paris 7, Irfu, Service d'Astrophysique, 
           CEA Saclay, F-91191 Gif-sur-Yvette, France\label{inst1}
      \and McWilliams Center for Cosmology, Department of Physics, Carnegie Mellon University, 
           Pittsburgh, PA 15213, USA\label{inst2}
      \and Department of Physics and Astronomy, University College London, Gower Street, 
           London WC1E 6BT, UK\label{inst3}
      \and Institut d'Astrophysique de Paris, CNRS UMR 7095 \& UPMC, 98 bis, boulevard
           Arago, 75014 Paris, France\label{inst4}}

\begin{document}

\abstract{Peak statistics in weak lensing maps access the non-Gaussian information contained 
in the large-scale distribution of matter in the Universe. They are therefore a promising 
complementary probe to two-point and higher-order statistics to constrain our cosmological models. 
Next-generation galaxy surveys, with their advanced optics and large areas, will measure 
the cosmic weak lensing signal with unprecedented precision. 
To prepare for these anticipated data sets, we assess the constraining power of peak counts in a 
simulated \textit{Euclid}-like survey on the cosmological parameters $\Omega_\mathrm{m}$, 
$\sigma_8$, and $w_0^\mathrm{de}$.
In particular, we study how $\textsc{Camelus}$---a fast stochastic model for predicting
peaks---can be applied to such large surveys. The algorithm avoids the 
need for time-costly $N$-body simulations, and its stochastic approach provides full PDF 
information of observables.
Considering peaks with signal-to-noise $\geq 1$, we measure the abundance histogram in a mock 
shear catalogue of approximately 5\,000 deg${}^2$ using a multiscale mass map filtering technique. 
We constrain the parameters of the mock survey using \textsc{Camelus} combined with 
approximate Bayesian computation, a robust likelihood-free inference algorithm.
Peak statistics yield a tight but significantly biased constraint in the $\sigma_8$--$\Omega_\mathrm{m}$ plane, 
as measured by the width $\Delta\Sigma_8$ of the 1-$\sigma$ contour. 
We find $\Sigma_8=\sigma_8(\Omega_\mathrm{m}/0.27)^\alpha=0.77_{-0.05}^{+0.06}$
with $\alpha=0.75$ for a flat $\Lambda$CDM model.
The strong bias
indicates the need to better understand and control the model's systematics before 
applying it to a real survey of this size or larger. 
We perform a calibration of the model and compare results to those from the two-point 
correlation functions $\xi_\pm$ measured on the same field. We calibrate the $\xi_\pm$ result 
as well, since its contours are also biased, although not as severely as for peaks.
In this case, we find for peaks $\Sigma_8=0.76_{-0.03}^{+0.02}$ with $\alpha=0.65$, while for
the combined $\xi_+$ and $\xi_-$ statistics the values are $\Sigma_8=0.76_{-0.01}^{+0.02}$ and $\alpha=0.70$.
We conclude that the constraining power can therefore be comparable between the two weak lensing 
observables in large-field surveys. Furthermore, the tilt in the $\sigma_8$--$\Omega_\mathrm{m}$
degeneracy direction for peaks with respect to that of $\xi_\pm$ suggests that a combined analysis 
would yield tighter constraints than either measure alone.
As expected, $w_0^\mathrm{de}$ cannot be well constrained without a tomographic 
analysis, but its degeneracy directions with the other two varied parameters are still 
clear for both peaks and $\xi_\pm$.}

\maketitle

\section{Introduction}\label{sec:intro}
The observed statistical distribution of matter in the Universe serves as a powerful discriminator 
of cosmological models. Different relative contributions to the Universe's mass-energy content 
produce different expansion histories and different amplitudes of matter clustering.
Weak gravitational lensing (WL), via its ability to probe the large-scale structure, has become a 
primary tool to study the total matter distribution in the Universe, as well as the nature of 
dark matter and dark energy.

The WL signal
consists of tiny coherent distortions of galaxy shapes whose light rays have been bent by gravitational
fields of dense matter structures lying along the line of sight. In contrast to the dramatic distortions 
seen in strong lensing systems, the images of weakly lensed galaxies are sheared and magnified 
only at the percent level relative to their original shapes. Weak lensing is therefore an inherently
statistical probe that, given a sufficient density of background sources, provides a means to 
map the projected matter distribution across the sky. Furthermore, such \textit{mass maps} trace 
the total matter distribution in an unbiased way, as WL is insensitive to the dynamical relationship 
between dark matter halos and the luminous galaxies that occupy them.

Numerous analyses have now constrained cosmological parameters using WL data from both ground- and 
space-based galaxy surveys. Recent studies include 
the \textit{Hubble Space Telescope} Cosmic Evolution Survey (COSMOS; \cite{SHJ.etal.2010}),
the Canada-France-Hawaii Telescope Lensing Survey (CFHTLenS; \cite{HVWM.etal.2012, HGH.etal.2013, KFH.etal.2013}),
the Sloan Digital Sky Survey (SDSS; \cite{HEH.etal.2014}),
the Dark Energy Survey (DES; \cite{TDESC.2015, DESSV.WL.2016}),
and the Kilo Degree Survey (KiDS; \cite{KHH.etal.2015, HVH.etal.2016}). 
Future surveys like \textit{Euclid} \citep{LAA.etal.2011} and the Large Synoptic Survey Telescope 
(LSST; \cite{LSST.SC.2009}) will afford unprecedented precision in WL measurements with their 
large survey areas and advanced optics. 
Cosmic shear studies planned for missions with a deeper but smaller field-of-view include the Hyper 
Supreme-Cam of the Subaru Telescope (HSC; \cite{MKN.etal.2012}) and the Wide-Field Infrared 
Survey Telescope-Astrophysics Focused Telescope Assets (WFIRST-AFTA; \cite{SGB.etal.2013}).

The basic WL measurements are the shear two-point correlation function (2PCF) and its 
associated power spectrum. These measures capture the second-order statistics, but as the shear 
distribution is not merely Gaussian, they neglect the non-Gaussian information encoded by the 
nonlinear formation of structure. Studying higher-order statistics of the shear and convergence
fields has therefore also become common. Examples of these include three-point correlation functions 
\citep{SSVW.etal.2011, FKE.etal.2014}, as well as $n$th order moments of the convergence field
and Minkowski functionals \citep{PHH.etal.2013, PLH.etal.2015}. 
The statistics of peaks, which are local maxima in WL maps, provide another way to access the 
non-Gaussian part of the signal.

Large signal-to-noise (S/N) peaks trace high-mass regions of the Universe and can often be
associated with massive galaxy clusters. High peaks therefore provide a direct and robust probe 
of the halo mass function. However, the origin of a WL peak does not typically admit of a unique 
interpretation. Low S/N peaks can arise from projection effects of the 
large-scale structure, or they can simply be spurious noise fluctuations. 
Studies aimed at \textit{detecting} massive galaxy clusters with WL thus focus on high peaks
with S/N larger than $\sim$4, whereas cosmological studies include the low peaks, since 
they also contain significant cosmological information \citep{YKH.etal.2013, LKP.2016}.

\cite{JvW.2000} initiated the study of WL peaks in their own right as a cosmological probe, 
using simulations to show that peaks can discriminate models with different total matter density 
parameters. 
Since then, many other papers have explored the efficacy of peaks for constraining 
both standard and nonstandard cosmological models
\citep{MSB.etal.2009, DH.2010, MAP.etal.2010, MFM.2011, MHS.etal.2011, PLS.2012, MSH.etal.2012, 
CCM.etal.2013, MSH.etal.2013, LK.2015a, LK.2015b, LKP.2016}.
Regarding recent surveys, peak analyses have also been used to derive constraints on 
$(\sigma_8,\Omega_\mathrm{m})$ with CFHTLenS data \citep{LPH.etal.2015}, CFHT Stripe-82 data 
\citep{LPL.etal.2015}, and DES Science Verification data \citep{KKF.etal.2016}.
\textit{Euclid} will observe approximately 15,000 deg${}^2$ of the extragalactic sky, an area
about two orders of magnitude larger than CFHTLenS and the currently available DES SV data. 
It is therefore important to study not only the statistical improvements that will be afforded 
by such a large survey, but also to anticipate potential challenges regarding systematics and
biases in our modeling of observations.

Our goal in this paper is to assess the ability of peak counts, modeled by the fast stochastic
algorithm \textsc{Camelus}, to constrain cosmological parameters in a large-area survey. 
\textsc{Camelus} was introduced in \cite{LK.2015a} and studied further in the context of parameter 
constraint strategies and mass map filtering methods in \cite{LK.2015b} and \cite{LKP.2016}, respectively.
See also \cite{ZMHH.etal.2016} for a recent comparison of \textsc{Camelus} predictions to 
$N$-body simulations over a broad range of cosmologies. 
The code uses a forward-model approach to generate lensing catalogues in a way that avoids 
time-costly $N$-body simulations and only requires a halo mass function and halo profile as input.

In this work, we apply the \textsc{Camelus} model to mock lensing data of area $\sim$5\,000 deg${}^2$.
Implementing a wavelet-based mass map filtering technique, we compute peak histograms of the mock 
catalogue as a function of S/N and filtering scale. The multiscale approach allows us build a peaks 
summary statistic, which separates out the cosmological information contained at different scales. 
Using \textsc{Camelus} with approximate Bayesian computation (ABC) for inference, we derive 
credible contours in $\sigma_8$--$\Omega_\mathrm{m}$ space and compute the derived parameter 
$\Sigma_8=\sigma_8(\Omega_\mathrm{m}/0.27)^\alpha$ for the mock survey. 
To compare with probes of Gaussian information, we compare contours and the uncertainty in 
$\Sigma_8$ from peaks to that of the two components of the 2PCF $\xi_+$ and $\xi_-$ of the 
shear field.

The remainder of the paper is organized as follows. In Sect. \ref{sec:theory}, we present the
basic theory of weak gravitational lensing relevant to this work. In Sect. \ref{sec:method}, 
we describe our methodology, including our multiscale mass-mapping technique,
the simulated galaxy catalogue used as observations, the \textsc{Camelus} model,
and finally our parameter inferences using ABC. Parameter 
constraint contours, as well as the comparison between peaks and $\xi_\pm$, are presented in Sect. 
\ref{sec:results}. We conclude in Sect. \ref{sec:conclusion}.

\section{Theoretical background}\label{sec:theory}
\subsection{Weak gravitational lensing}
Throughout this work, we assume that the Universe can be described by a weakly perturbed FLRW model 
with Newtonian potential $\Phi$. Local deviations from the average matter density $\bar{\rho}$ are 
characterized by the density contrast,
\begin{equation}
   \delta(t,\bm{\theta},w) = \frac{\rho(t,\bm{\theta},w)-\bar{\rho}(t)}{\bar{\rho}(t)},
\end{equation}
where coordinates $(t, \bm{\theta}, w)$ represent cosmic time, angular position, and comoving 
radial distance, respectively. Density perturbations are related to $\Phi$ via Poisson's equation 
for a pressureless ideal fluid
\begin{equation}
   \nabla^2 \Phi = 4\pi\mathrm{G}a^2\bar{\rho}\delta,
\end{equation}
which can be written more conveniently in terms of cosmological parameters as
\begin{equation}
   \nabla^2 \Phi = \frac{3}{2}H_0{}^2\Omega_\mathrm{m}\frac{\delta}{a},
\end{equation}
where $H_0$ is the present Hubble parameter, $\Omega_\mathrm{m}$ is the present matter density, and $a(t)$ 
is the scale factor of the Universe. 

The images of distant galaxies undergo distortions as their light propagates through regions of non-uniform 
potential on their way to us. The original unlensed positions $\bm{\beta}$ of light beams undergo a remapping 
to the positions $\bm{\theta}$ at which we observe them, which we can quantify by the Jacobian matrix 
$\mathcal{A}_{ij}=\partial\beta_i/\partial\theta_j$ of the transformation. It can be shown that to linear 
order in $\Phi$, $\mathcal{A}$ is determined by a line-of-sight integral of second-order transverse 
derivatives of $\Phi$; see \cite{SVWJ.etal.1998}, for example. This motivates the definition of the lensing 
potential
\begin{equation}
   \psi(\bm{\theta},w) = \frac{2}{c^2}\int_0^w~\mathrm{d}w'~\frac{f_K(w-w')}{f_K(w)f_K(w')}\Phi(f_K(w')\bm{\theta}, w')
\label{eq:psi}
\end{equation}
and the decomposition of $\mathcal{A}$ in terms of the convergence $\kappa(\bm{\theta})$ and shear 
$\gamma(\bm{\theta}) = \gamma_1(\bm{\theta}) + \mathrm{i}\gamma_2(\bm{\theta})$:
\begin{equation}
   \mathcal{A} = 
     \begin{pmatrix}
       1 - \kappa - \gamma_1 & -\gamma_2 \\
       -\gamma_2 & 1 - \kappa + \gamma_1
     \end{pmatrix}.
\end{equation}
In the above, $f_K(w)$ is the comoving angular diameter distance
\begin{equation}
   f_K(w) = 
   \begin{cases}
     ~\frac{1}{\displaystyle\sqrt{K}}\sin\left(\sqrt{K}~w\right)    \hfill\qquad \mathrm{if} & K > 0, \\[1ex]
     ~w                                                             \hfill\qquad \mathrm{if} & K = 0, \\[1ex]
     ~\frac{1}{\displaystyle\sqrt{-K}}\sinh\left(\sqrt{-K}~w\right) \hfill\qquad \mathrm{if} & K < 0~,
   \end{cases}
\end{equation}
and $K$ is the curvature of space. Then $\kappa$ and $\gamma$ are expressible directly as second-order
derivatives of the lensing potential
\begin{equation}
  \kappa = \frac{1}{2}\left(\partial_1{}^2 + \partial_2{}^2\right)\psi,
\label{eq:kappa}
\end{equation}
\begin{equation}
  \gamma_1 = \frac{1}{2}\left(\partial_1{}^2 - \partial_2{}^2\right)\psi
  \qquad \mathrm{and} \qquad \gamma_2 = \partial_1\partial_2\psi,
\label{eq:gamma}
\end{equation}
where $\partial_i$ denotes the partial derivative with respect to $\theta_i$. In particular, we can now
relate $\kappa$ directly to the density fluctuations as
\begin{equation}
  \kappa(\bm{\theta}, w) = \frac{3}{2}\frac{H_0{}^2\Omega_\mathrm{m}}{c^2}\int_0^w~\mathrm{d}w'~\frac{f_K(w')f_K(w-w')}{f_K(w)}\frac{\delta(f_K(w')\bm{\theta}, w')}{a(w')},
\label{eq:kappa_integral}
\end{equation}
making the interpretation of convergence as the projected mass density more apparent.

The explicit time dependence of $\Phi$ in Eqs. (\ref{eq:psi}) and (\ref{eq:kappa_integral}) has been suppressed, 
since the integral over $w'$ is understood as being performed along our past light cone. Further, we 
restrict our attention to the weak lensing regime of small deviations 
(i.e., where $|\kappa|$, $|\gamma| \ll 1$), and the Born approximation 
permits us to integrate along the unperturbed light path.

Further details of weak lensing theory and formalism can be found in the reviews of, for example,
\cite{BS.2001}, \cite{HJ.2008}, and \cite{Kilbinger.2015}.

\subsection{Mass mapping}
Mass maps refer to maps of the convergence $\kappa$, a quantity which directly represents the matter distribution
projected along the line of sight. This is readily seen in the integral of Eq. (\ref{eq:kappa_integral}).
As they are scalar fields, mass maps are more convenient for many applications compared to the 
complex shear $\gamma$, as in, for example, cross-correlating with other scalar fields like the 
galaxy distribution or cosmic microwave background (CMB) temperature. In this work, we use mass 
maps to facilitate the straightforward identification of weak lensing peaks.

In practice, we cannot directly measure $\kappa$ or $\gamma$ from a galaxy survey. Instead what we 
observe are galaxy ellipticities, which are an additive combination of intrinsic ellipticity and the 
shear due to lensing. Assuming that the source galaxies are not intrinsically aligned, we can 
therefore average the observed ellipticities over many sources in a small region of the sky to 
estimate the shear signal $\gamma$. This approximation works in the weak lensing regime, since 
$\epsilon^\mathrm{obs} \approx \epsilon^\mathrm{int} + g$,  where $g = \gamma/(1-\kappa)$ is the reduced
shear. The average of many observed ellipticities is thus an accurate measure of $g\approx\gamma$.

The inversion method of Kaiser and Squires \citep{KS.1993} gives a parameter-free prescription 
for computing $\kappa$ from $\gamma$ via the Fourier transforms of Eqs. (\ref{eq:kappa}) and 
(\ref{eq:gamma}). The equations become
\begin{equation}
  \tilde{\kappa}(\bm{\ell}) = \tilde{D}^*(\bm{\ell})\tilde{\gamma}(\bm{\ell}),
\end{equation}
where $\tilde{\kappa}$ and $\tilde{\gamma}$ are the Fourier transforms of $\kappa$ and $\gamma$, 
$\bm{\ell}$ is the Fourier counterpart to angular position $\bm{\theta}$, ${}^*$ denotes complex 
conjugation, and 
\begin{equation}
  \tilde{D}(\bm{\ell}) = \frac{\ell_1^2 - \ell_2^2 + 2\mathrm{i}\ell_1\ell_2}{|\bm{\ell}|^2}
  \qquad\mathrm{for}\quad \bm{\ell}\neq\bm{0}.
\end{equation}
Convergence can therefore be determined directly from measurements of galaxy shapes in the weak
lensing regime, up to an additive constant. 
It is well known that Kaiser-Squires inversion creates undesirable artifacts at the 
boundaries of finite fields. We address this issue by excluding a border of 8 pixels 
($=4$ arcmin) around each 5 x 5 deg${}^2$ convergence map generated throughout our analysis.

\subsection{Shear two-point correlation functions}
A standard approach to constraining cosmology with weak lensing analyses is to use $N$-point 
correlation functions of the shear field. Primary among these are the two-point functions
$\xi_+(\bm{\theta})$ and $\xi_-(\bm{\theta})$. We present a brief overview of the $\xi_\pm$
statistics, since we later compare their parameter constraint results to those of peak counts.

The shear $\gamma$ at angular position $\bm{\theta}$ can be decomposed into a 
tangential component $\gamma_\mathrm{t}$ and cross component $\gamma_\times$ with respect to a 
point at position $\bm{\theta}_0$. If the separation vector 
$\bm{\theta}-\bm{\theta}_0$ has polar angle $\varphi$, then
\begin{equation}
  \gamma_\mathrm{t} = -\mathrm{Re}[\gamma\,\exp(-2\mathrm{i}\varphi)], \qquad
  \gamma_\times = -\mathrm{Im}[\gamma\,\exp(-2\mathrm{i}\varphi)].
  \label{eq:gamma_tx}
\end{equation}
The minus sign ensures that $\gamma_\mathrm{t} > 0$ for tangential alignment with respect to the
position $\bm{\theta}_0$, and $\gamma_\mathrm{t} < 0$ for radial alignment.

The shear 2PCF $\xi_\pm$ are defined as combinations of $\gamma_\mathrm{t}$
and $\gamma_\times$ as \citep{Miralda-Escude.1991}
\begin{equation}
  \xi_\pm(\theta) = \langle \gamma_\mathrm{t}\gamma_\mathrm{t} \rangle (\theta) \pm
                    \langle \gamma_\times\gamma_\mathrm\times \rangle (\theta)
\end{equation}
as a function of angular scale $\theta$.
Following \cite{SVWM.2002}, we can estimate $\xi_\pm$ from pairs of measured galaxy ellipticities as
\begin{equation}
  \hat{\xi}_\pm(\theta) = \frac{\sum_{ij} w_i w_j \left[\epsilon_\mathrm{t}(\bm{\theta}_i)\epsilon_\mathrm{t}(\bm{\theta}_j) \pm
                          \epsilon_\times(\bm{\theta}_i)\epsilon_\times(\bm{\theta}_j)\right]}{\sum_{ij} w_i w_j}.
  \label{eq:xi_pm}
\end{equation}
Here $\epsilon_\mathrm{t}$ and $\epsilon_\times$ are defined analogously as $\gamma_\mathrm{t}$
and $\gamma_\times$ in Eq. (\ref{eq:gamma_tx}). Weights are denoted by $w$, and the indices $i,j$ 
refer to galaxies at positions $\bm{\theta}_i$ and $\bm{\theta}_j$. The summation is carried out 
over all galaxy pairs with $|\bm{\theta}_i - \bm{\theta}_j|$ lying within an angular bin around $\theta$.

\section{Methodology}\label{sec:method}
Our goal is to study the constraints on cosmological parameters attainable from peak count 
statistics in a wide-field survey like \textit{Euclid}. We describe in
the following sections the simulated \textit{Euclid}-like galaxy catalogue we have chosen for 
analysis, as well as the algorithm we use to simulate stochastic lensing catalogues
as a function of cosmological parameters. We then describe the multiscale mass-mapping technique
we use to construct peak abundance data vectors for both the mock catalogue and for those simulated 
by \textsc{Camelus}. Finally, we present the method of approximate Bayesian computation, 
which we use for parameter inference.

\subsection{Mock Observations}\label{ss:MICE}
The Marenostrum Institut de Ci{\`e}ncies de l'Espai Grand Challenge simulation
\citep{FGC.etal.2008, CFC.etal.2010, MICE.I.2015, MICE.II.2015, MICE.III.2015, CCG.etal.2015, 
HBG.etal.2015}
is a set of large-volume $N$-body runs carried out at the Barcelona Supercomputing Center using the 
\textsc{GADGET-2} $N$-body code \citep{Springel.2005}. It contains approximately 70 billion dark 
matter particles in a simulation box of comoving side length $\sim3 h^{-1}$ Gpc, giving high mass 
resolution in a large and statistically independent (i.e., non-repeated) volume. Using a technique 
combining halo occupation distribution (HOD) and halo abundance matching (HAM) to populate dark 
matter halos, a mock galaxy catalogue was generated with a light cone extending out to $z=1.4$. 
The MICECATv2.0 catalogue\footnote{\url{http://cosmohub.pic.es/}} 
(MICE hereafter) we have used in this work covers an area 
of about 5\,000 deg${}^2$---a contiguous octant of the sky lying within 
$0^\circ \le \textrm{RA} \le 90^\circ$ and $0^\circ \le \textrm{Dec} \le 90^\circ$---which is
approximately one-third the size of what is expected for Euclid's wide-field survey.
The catalogue contains lensing information for nearly 500 million magnitude-limited galaxies,
with completeness depending on redshift and position. In the least complete areas, for example, 
the catalogue is complete to observation band $H\approx 24\,(23)$ up to $z\approx 0.45\,(1.4)$. In
this work, we only make use of the galaxy positions, observed ellipticities, and true redshifts.

\begin{figure}
\resizebox{\hsize}{!}{\includegraphics{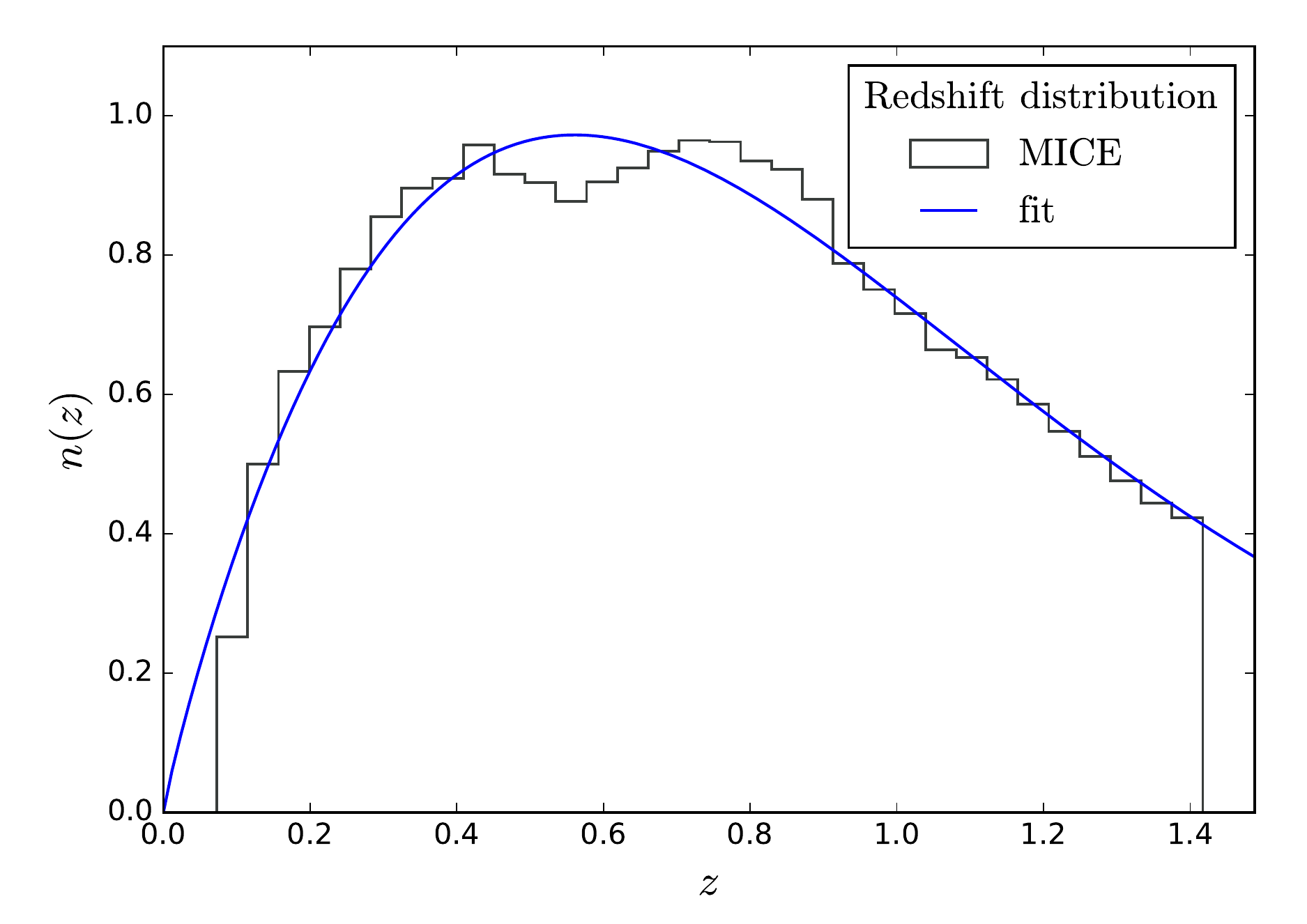}}
\caption{Histogram of the true redshift distribution of the MICE mock catalogue. The
         solid (blue) line is the best-fit $n(z)$ in the form of Eq. (\ref{eq:nz}) with parameters
         $(p,q,z_0)=(0.88,1.40,0.78)$. The normalization is computed numerically.}
\label{fig:z_dist}
\end{figure}

The MICE cosmology is a flat $\Lambda$CDM model with parameters $\Omega_\mathrm{m}=0.25$, 
$\Omega_\Lambda=0.75$, $\Omega_\mathrm{b}=0.044$, $\sigma_8=0.8$, $h=0.7$, and $n_s=0.95$.
We find that its galaxy redshift distribution is well approximated by the parameterized form
\begin{equation}
  n(z) \propto \left(\frac{z}{z_0}\right)^p~
         \exp\left[-\left(\frac{z}{z_0}\right)^q\right],
\label{eq:nz}
\end{equation}
with best-fit parameters $p=0.88$, $q=1.40$, and $z_0=0.78$ based on chi-square 
minimization. The normalization factor is determined by numerical integration.
Figure \ref{fig:z_dist} shows the agreement between $n(z)$ and the histogram of true 
redshifts from MICE. We use this functional form as input to the \textsc{Camelus} algorithm in order 
to generate its simulated catalogues. With real \textit{Euclid} data, we would instead use the
estimated photo-$z$ distribution, as true galaxy redshifts are not known.

We assign intrinsic ellipticities to the MICE galaxies so that the distributions of the two components 
$\epsilon_1$ and $\epsilon_2$ match those observed in the COSMOS survey. These closely approximate 
Gaussian distributions with zero mean and a standard deviation of 0.3. We generate observed 
ellipticities, which are used for mass mapping, by the relations \citep{SS.1997}
\begin{equation}
  \epsilon_\mathrm{obs} = 
   \begin{cases}
     ~ \displaystyle{\frac{\epsilon_\mathrm{int} + g}{1 + g^*\,\epsilon_\mathrm{int}}} \hfill\qquad \mathrm{if} & |g| \leq 1, \\[3ex]
     ~ \displaystyle{\frac{1 + g\,\epsilon_\mathrm{int}^*}{\epsilon_\mathrm{int}^* + g^*}} \hfill\qquad \mathrm{if} & |g| > 1,
   \end{cases}
\end{equation}
where $\epsilon_\mathrm{obs}$ and $\epsilon_\mathrm{int}$ are the observed and intrinsic 
ellipticities, respectively, and ${}^*$ represents complex conjugation. Shear $\gamma$ and 
convergence $\kappa$ values that comprise $g$ are provided in the MICE catalogue for each galaxy 
line of sight. Given that $|g| > 1$ for only a few of galaxies, the resulting standard deviations 
of the observed ellipticity components remain at 0.3.

The \textsc{Camelus} algorithm, which we describe in Sect.\ref{ss:CAMELUS}, produces shear 
catalogues of size 25 deg${}^2$ in Cartesian space. For consistency, we therefore also extract 
and analyze 25 deg${}^2$ patches from the full MICE catalogue. We aim for as many independent
patches as possible to ensure the best statistical constraining power of the peak count information.
To do this, we take the straightforward approach of dividing the MICE octant into strips along 
lines of constant declination (Dec) such that each strip spans at least 5 deg in Dec. We then 
divide each strip into rectangles along lines of constant right ascension (RA) such that after a 
gnomonic (flat-sky) projection, we achieve the maximum number of 25 deg${}^2$ squares for the 
strip. 

The transformation equations are
\begin{equation}
  x = \frac{\cos\phi \sin(\lambda-\lambda_0)}{\cos\phi_0 \cos\phi \cos(\lambda-\lambda_0)
      + \sin\phi_0 \sin\phi}
\end{equation}
and
\begin{equation}
  y = \frac{\cos\phi_0 \sin\phi - \sin\phi_0 \cos\phi \cos(\lambda-\lambda_0)}{\cos\phi_0 
       \cos\phi \cos(\lambda-\lambda_0) + \sin\phi_0 \sin\phi},
\end{equation}
where $(\lambda,\phi)$ are the (RA, Dec) coordinates to be projected, $(\lambda_0,\phi_0)$
is the projection center, and $(x, y)$ are the tangent plane coordinates.
The non-conformal gnomonic projection has the useful property of mapping great circles to
straight lines, but it introduces shape and distance distortions in the tangent plane radially away 
from the origin. For the small size of the patches we consider, however, these effects are 
sub-percent level. For example, the distortion error in both the area and the maximum angular 
separation within the 25 deg${}^2$ fields is $< 0.3\%$.

Due to the spherical geometry, the number of usable patches within each strip varies significantly 
with declination. The strip with its edge at the equator contains 17, and the one nearest the
north pole contains only one.

\begin{figure}
\resizebox{\hsize}{!}{\includegraphics{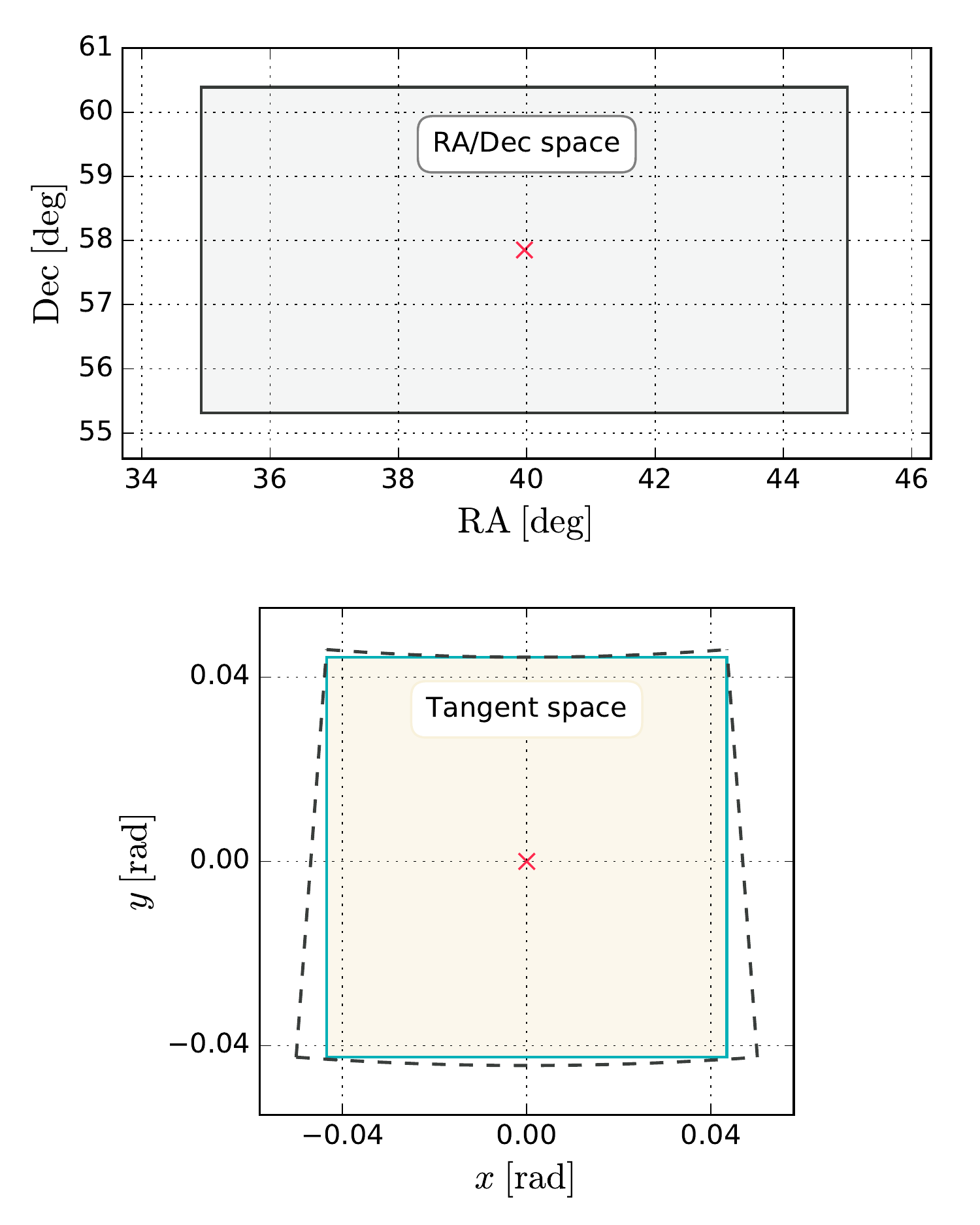}}
\caption{Example of our patch extraction and projection scheme for MICE. The upper panel shows the 
         original extent of the patch, which lies within $34.9^\circ \leq \mathrm{RA} \leq 45.0^\circ$, 
         $55.3^\circ \leq \mathrm{Dec} \leq 60.4^\circ$. This area is projected into the region enclosed
         by the dashed line in the lower panel of the tangent space. The bounded inner square
         represents the 25 deg${}^2$ area for which we generate the multiscale maps and compute 
         peak histograms.}
\label{fig:projection}
\end{figure}

Figure \ref{fig:projection} shows an example of the extraction and projection geometry for a patch
centered on $(\mathrm{RA},\mathrm{Dec})=(40.0, 57.9)$ deg. In the upper panel, the shaded rectangular 
region is the area in RA/Dec space before projection. This area is projected about the origin,
marked by a red X, into the region enclosed by the dashed line in the lower panel. The
inner tan square in the tangent space represents the area we use for mass-mapping for this patch.

This cutting is clearly not optimal in the sense that some areas are ultimately excluded from our 
analysis, namely those between the dashed and solid lines of the lower panel. All of the galaxies 
in the shaded RA/Dec rectangle end up inside the dashed region, but only those that fall in the 
tan square are used. Despite the simple approach, we still achieve $186 \times 25 = 4\,650$ 
deg${}^2$ of total effective area for our analysis. This should be sufficient to guide our 
intuitions about the constraining power of peak statistics from a survey like \textit{Euclid}.

\subsection{Modeling peak counts with fast simulations}\label{ss:CAMELUS}
In this section we briefly describe the \textsc{Camelus} algorithm that we use to generate shear 
catalogues from dark matter halo simulations. 
The code has been developed and tested in \cite{LK.2015a,LK.2015b} and \cite{LKP.2016}, and
the software is available on GitHub\footnote{\url{http:github.com/Linc-tw/camelus}}.
We refer the interested reader to those papers for a more complete description. 

The first step is to sample halo masses from a mass function, which is taken to have the form 
\citep{JFW.etal.2001}
\begin{equation}
  f(\sigma) = 0.315\exp\left(-|\ln \sigma^{-1} + 0.61|^{3.8}\right).
\end{equation}
The parameter $\sigma(z,M)$, which serves as a proxy for mass, is defined as the standard deviation 
of the linear density field, which has been smoothed with a spherical top-hat filter of radius $R$ 
such that $M=(4\pi/3)\bar{\rho}_0 R^3$. Halos are selected with masses within the range
$5\cdot10^{12}\,h^{-1}M_{\odot} \leq M \leq 10^{17}\,h^{-1}M_{\odot}$.

Next, the sampled halos are distributed randomly in an observation field of $5\times5$ deg${}^2$. 
The radial mass distributions of each halo are taken to be that of a truncated Navarro-Frenk-White 
(NFW) profile 
\citep{NFW.1996,NFW.1997}
\begin{equation}
  \rho(r) = \frac{\rho_s}{(r/r_s)^{\alpha_\mathrm{NFW}}(1 + r/r_s)^{3-\alpha_\mathrm{NFW}}}\Theta(r_\mathrm{vir}-r),
\end{equation}
where $\rho_s$ is the characteristic overdensity related to the halo's concentration, 
$r_s$ is the scale radius, $r_\mathrm{vir}$ is the virial
radius, $\alpha_\mathrm{NFW}$ is the inner slope, and $\Theta$ is the Heaviside step function. The scale radius
can be written as $r_s=r_\mathrm{vir}/c$, the ratio of the physical virial radius to the concentration
parameter. Following \cite{TJ.2002}, we assume a parameterized form of $c$ as a function of redshift 
and mass,
\begin{equation}
  c(z,M) = \frac{c_0}{1+z}\left(\frac{M}{M_\star}\right)^{-\beta}.
\end{equation}
Here, the pivot mass $M_\star$ is defined such that $\sigma(M_\star)$ equals the critical density for 
spherical collapse at $z=0$. We fix the parameters as $(c_0,\beta)=(9,0.13)$, as we find these to 
give good agreement with the MICE data in terms peak count histograms across different filtering scales.
These, as well as other parameter settings adopted in this work, are summarized in Table 
\ref{table:settings}. 

\begin{table}
\caption{Parameter settings for \textsc{Camelus}}
\label{table:settings}
\centering
\def\arraystretch{1.15}
\begin{tabular}{c c c}
\hline\hline
Description & Symbol & Value \\
\hline
  Dimensionless Hubble parameter    & $h_{100}$             & 0.7   \\
  Baryon density                    & $\Omega_\mathrm{b}$   & 0.044 \\
  Spectral index                    & $n_s$                 & 0.95  \\
  DE linear EOS parameter           & $w_1$                 & 0.0   \\
\hline
  NFW inner slope                   & $\alpha_\mathrm{NFW}$ & 1.0   \\
  $M$--$c$ relation parameter       & $c_0$                 & 9.0   \\
  $M$--$c$ relation parameter       & $\beta$               & 0.13  \\
  Galaxy density [arcmin${}^{-2}$]  & $n_\mathrm{gal}$      & 27    \\
  Source ellipticity dispersion     & $\sigma_\epsilon$     & 0.43  \\
\hline
\end{tabular}
\end{table}

The final step is to generate a simulated catalogue of reduced shear values for source galaxies in the
field. Sources are distributed randomly according to Eq. (\ref{eq:nz}) and assigned ellipticities from
a Gaussian distribution with dispersion 
$\sigma_\epsilon=(\sigma_{\epsilon_1}^2+\sigma_{\epsilon_2}^2)^{1/2}=0.43$. Orientations are random, 
meaning that we neglect intrinsic alignment in this work.
The projected lensing quantities are then calculated, which can be done analytically for an NFW profile;
see \cite{TJ.2003a,TJ.2003b}, for example. Intrinsic source ellipticities are combined with the computed 
reduced shear $g$ to give the final simulated lensing catalogue. We make S/N maps from this catalogue 
and compute peak count statistics following the procedure described in Sect. \ref{ss:maps}.

\begin{figure*}
\centering
\includegraphics[width=17cm]{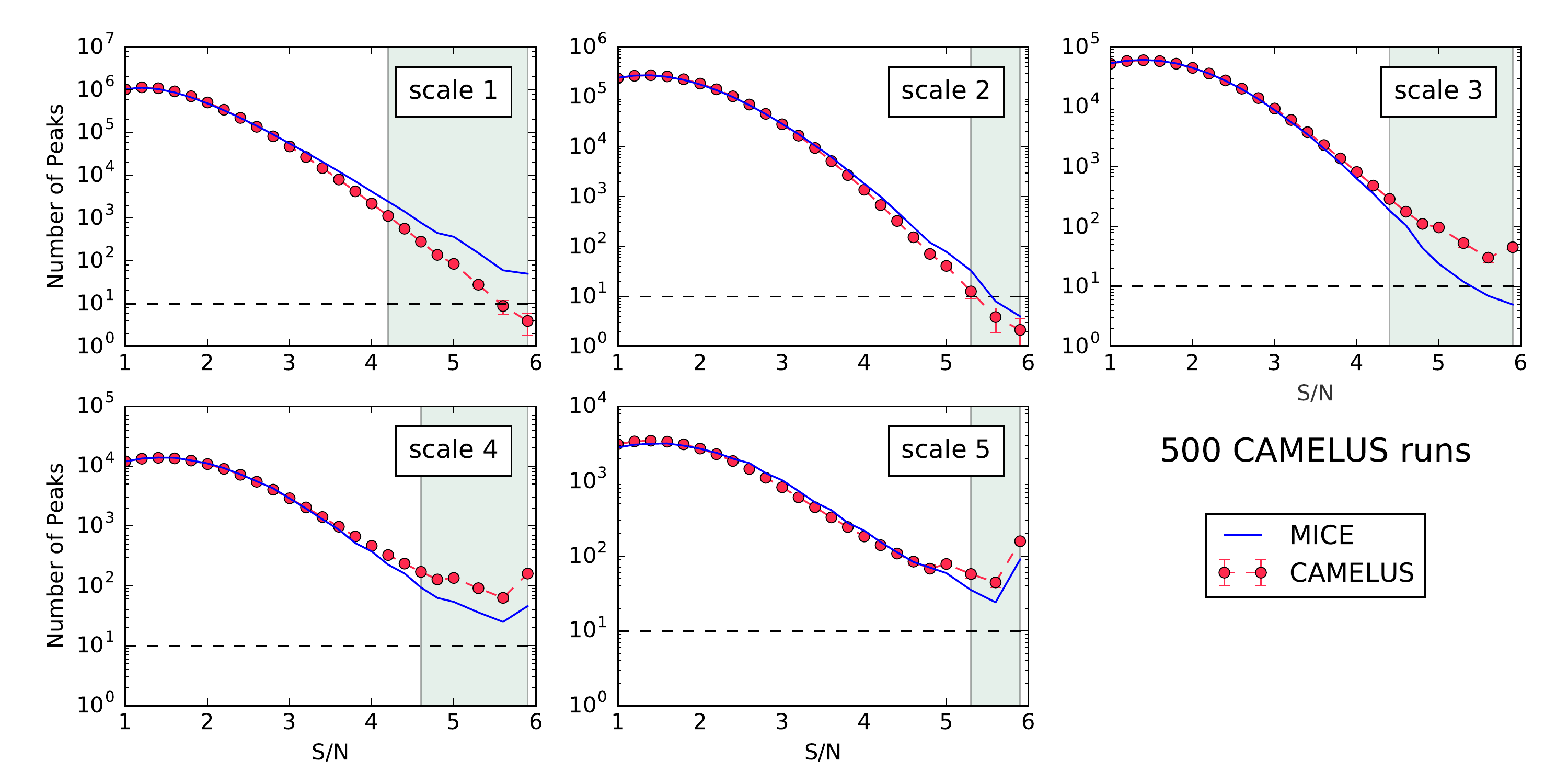}
\caption{Comparison of peak abundance histograms between \textsc{Camelus} (red circles) and MICE 
         (blue line) 
         for 4\,650 deg${}^2$. Error bars for \textsc{Camelus} represent the standard deviation 
         of 500 realizations for the associated S/N bin. There is good agreement between the mock 
         observations and the \textsc{Camelus} predictions across all wavelet filtering scales, 
         especially for $\nu$ less than $\sim$5. The shaded regions represent bins where 
         the relative different between MICE and \textsc{Camelus} is larger than 50\%. We apply 
         a conservative cut and omit these higher-bias bins in our analysis.}
\label{fig:histograms}
\end{figure*}

We note here some important assumptions implicit in the \textsc{Camelus} approach to simulating 
weak lensing maps. The first is that peak number counts arise primarily due to bound matter, and 
not from the diffuse matter that makes up, for example, cosmic filaments. The second is that the
spatial correlation of dark matter halos does not have a significant impact on peaks. The results 
of \cite{LK.2015a} support the reasonableness of these assumptions for relatively small surveys. 
The present work, however, reveals limitations of \textsc{Camelus} in the context of larger survey 
areas; see Fig. \ref{fig:histograms} below and the parameter constraint results in Sect. 
\ref{sec:results}.

\begin{figure*}
\centering
\includegraphics[width=17cm]{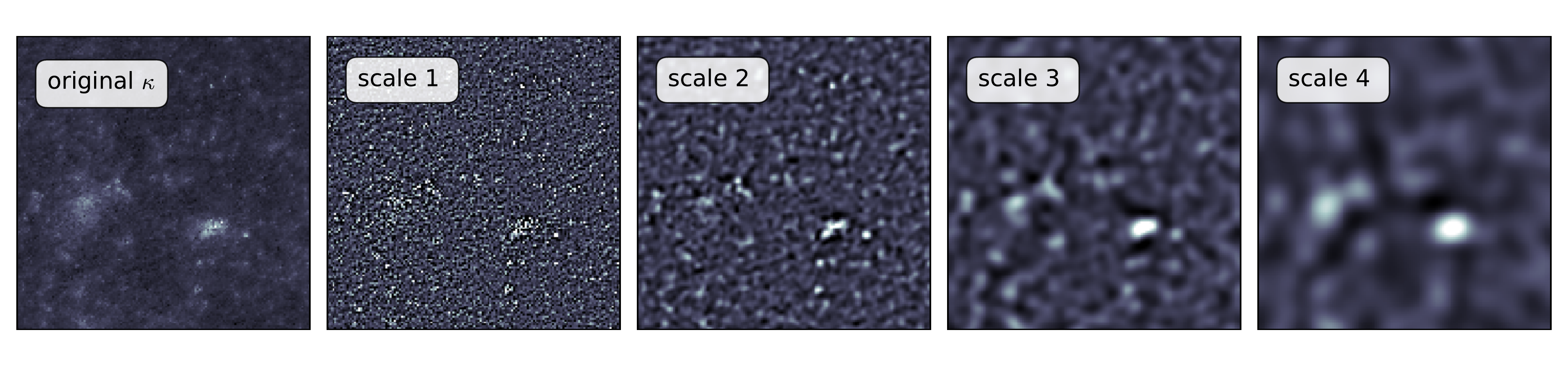}
\caption{Starlet decomposition of an example noiseless convergence map $\kappa(\bm{\theta})$.
         The images are $150\times150$ with pixels of size 0.5 arcmin. The original map is shown 
         along with the wavelet coefficient maps $\{w_j\}$ up to $j_\mathrm{max}=4$. The final 
         smoothed map $c_5$, i.e., the low-pass filtered version of $\kappa$, is not shown. 
         One can see that the transform picks out features of $\kappa$ at successively larger 
         scales as $j$ increases.}
\label{fig:kappa_mr}
\end{figure*}

\subsection{Multiscale wavelet filtering}\label{ss:maps}
Weak Lensing maps are dominated by galaxy shape noise and must be filtered to access their signal. 
Numerous schemes have been employed to de-noise WL maps, the most common of which is filtering with 
a Gaussian kernel. For the reasons described below, we choose to filter the mass maps in 
our analysis using the isotropic undecimated wavelet, or \textit{starlet}, transform 
\citep{SFM.2007}.

The starlet has many properties that make it useful in astrophysical image processing. 
First, isotropy makes it well-suited to extract features from astrophysical data containing 
objects that are roughly round, such as stars, galaxies, and clusters. Next, it is a multiscale
transform in which the information contained at different scales in a image is 
separated out simultaneously.
The filter functions associated with the starlet transform are localized in real 
space---i.e., they go to zero within a finite radius. The first wavelet function acts as a 
high-pass filter, while the remaining wavelets act as band-pass filters for their respective 
scales, since they are localized in Fourier space as well. 
Finally, the wavelet functions are compensated, meaning they integrate to zero over their 
domains. This is beneficial, as it was shown by \cite{LKP.2016} that in the context of peak-count 
analyses, compensated filters are better at capturing cosmological information compared to 
non-compensated filters like the Gaussian kernel. We note that the wavelet transform of a 
convergence map at a particular scale is also formally equivalent to aperture-mass filtering by
a corresponding compensated filter \citep{LPS.2012}.
%

The starlet transform of an $N\times N$ image $I$ amounts to successive convolutions of 
the image with a set of discrete filters corresponding to different resolution scales 
$j=1,...,j_\mathrm{max}$.
The result is a set of $J=j_\mathrm{max}+1$ maps $\{w_1,...,w_{j_\mathrm{max}},c_J\}$, 
each of size $N\times N$, where the `detail' maps $\{w_j\}$ represent $I$ filtered at a scale of 
$2^j$ pixels. The final map $c_J$ represents a smoothed version of $I$, and
the original image is exactly recoverable from the decomposition: 
$I = c_J + \sum_{j=1}^{j_\mathrm{max}}w_j$. Further details, including explicit expressions for 
the discrete filter bank associated with the starlet can be found in \cite{SFM.2007}.

As an illustration, we show the starlet transform of a convergence map of $150\times 150$ 
pixels in Fig. \ref{fig:kappa_mr}. The original image is a noiseless map $\kappa(\bm{\theta})$ with
pixels of size 0.5 arcmin, and the decomposition is shown for $j_\mathrm{max}=4$ wavelet scales 
increasing to the right. Brighter pixels indicate larger values. The smoothed map $c_5$, which is 
the low-pass filtered version of $\kappa$, is not shown. One can see that progressively 
lower frequency features are picked out from $\kappa$ as $j$ increases.

After filtering, the noise level is different at different scales, and these levels are 
related by the ratio of the filter norms. If we denote the wavelet convolution kernel at scale 
$j$ by $W_j$, then the noise at this scale is
\begin{equation}
  \sigma_j^2 = \sigma_\mathrm{ref}^2\cdot\frac{||W_j||_2^2}{||W_\mathrm{ref}||_2^2},
\end{equation}
where $||\cdot||_2$ denotes the $\ell_2$ norm, and $\sigma_\mathrm{ref}$ is the known 
noise level at a particular reference scale.

Instead of using the analytic expression corresponding to the input galaxy ellipticity dispersion,
we estimate $\sigma_\mathrm{ref}$ from the data as follows. 
Let $\kappa_j$ denote the map filtered at scale $j$ (analogous to $w_j$ above).
At the finest wavelet scale, $j=1$, the noise is dominant, and we expect the wavelet coefficients 
to follow the noise distribution.
We therefore take $\sigma_\mathrm{ref}$ as the dispersion of $\kappa_1$, the smallest resolution 
of the convergence map decomposition.

The signal-to-noise at scale $j$ of a given coefficient at location $\bm{\theta}$ can then be 
written as
\begin{equation}
  \nu_j(\bm{\theta})=\frac{\kappa_j(\bm{\theta})}{\sigma_j},
\end{equation}
where we only consider positive values of $\kappa_j$.
For all maps used in our cosmological analysis, we bin the galaxies into pixels of size 0.5 arcmin, 
resulting in maps that are $600\times600$ pixels. We transform $\kappa$ with 
$j_\mathrm{max}=5$ wavelet scales and identify peaks in these maps as local maxima 
of $\nu$, where a peak is simply a pixel with a larger $\nu$ value than its eight neighboring pixels.

\subsection{Choice of data vector}\label{sec:data_vector}
To compare the predictions of \textsc{Camelus} to MICE, and ultimately to constrain parameters, 
we need a summary statistic that encapsulates the peak information contained at different wavelet
scales. For this, we choose peaks of $\nu \geq 1$ with bin boundaries defined by
$[1.0,1.2,...,4.8,5.0]\cup[5.3,5.6,5.9,+\infty)$. That is, the bin spacing is 
$\Delta\nu = 0.2$ for $\nu \in [1.0, 5.0]$ and $\Delta\nu = 0.3$ for $\nu \in [5.0, 5.9]$.

\cite{LKP.2016} have shown that keeping the multiscale peak histograms separate, rather than
combining them into a single data vector, yields tighter constrains on cosmological parameters.
Following this result, we adopt a summary statistic that is the concatenation of peak-count 
histograms at \textit{five} consecutive wavelet scales. The scales are arranged in order of 
decreasing resolution, analogous to the four scales in the example 
of Fig. \ref{fig:kappa_mr}.

In Fig. \ref{fig:histograms}, we show a comparison of the MICE peak histograms at each scale 
with the \textsc{Camelus} predictions for the 4\,650 deg${}^2$ field. MICE results are shown
as solid blue lines, \textsc{Camelus} as red circles. The data points for both align with the left
edges of their associated S/N bins. The \textsc{Camelus} data were obtained by averaging 500 
realizations of the code with the MICE cosmology as input, and error bars represent the standard 
deviations of these runs. 

The plots reveal overall good agreement between the mock observations and the \textsc{Camelus} 
predictions across all scales, especially for $\nu$ less than $\sim$5. 
The highest S/N bins exhibit the largest bias at each scale, and the shaded regions 
indicate bins where the relative difference between MICE and the \textsc{Camelus} prediction 
exceeds 50\%. Scales 2 and 5, corresponding to angular filtering scales of 2 and 16 arcmin, 
respectively, provide the best agreement in the high S/N range. Furthermore, \textsc{Camelus}
predicts systematically fewer peaks than MICE at the first two scales, while the reverse is true
for scales 3 and 4. This suggests that there could be an optimal filtering scale lying between 
scales 2 and 3; however, \textsc{Camelus} underpredicts the MICE peaks again at scale 5 over most
of the range, making the choice of an optimal scale ambiguous.

There are numerous competing factors that contribute to the difference between \textsc{Camelus} peak
predictions and MICE. The first is related to differences in the halo mass functions between the 
model and the data. Halo sampling by \textsc{Camelus} agrees well with the Jenkins model it is
based on \citep{LK.2015a}. On the other hand, the Jenkins model underpredicts the measured halo 
abundance of the MICE simulation substantially for $M > 10^{14}\,h^{-1}M_{\odot}$ 
\citep{CFC.etal.2010, MICE.II.2015}. One would therefore expect the proportion of high-mass halos 
in a given field area to be smaller for \textsc{Camelus} than for MICE. Further, as most high 
S/N peaks are due to single massive halos \citep{YKW.etal.2011}, the peak counts in the high 
S/N bins should be larger for MICE than for \textsc{Camelus}. We see in Fig. \ref{fig:histograms}
that this is only true for the first two wavelet scales.

Another reason for the differences is the lack of halo clustering in the \textsc{Camelus} model. 
It was shown in \cite{LK.2015a} that the effect of randomizing the angular positions
of halos is to reduce the peak count number density by between 10\% and 50\%. The reason is that
decorrelating halos leads to less overlap between them on the sky and therefore to a decrease
in the number of high peaks. This effect is offset, however, by the replacement of $N$-body halos
by spherical NFW profiles, as well as ignoring the contribution of unbound matter to the lensing
signal. A third consideration is that the S/N of a halo's WL signal will
be maximized when the filter size approximately matches that of the structure,
which depends on its specific parameters like mass, concentration, and redshift.

Interpreting the precise nature of the differences between MICE and \textsc{Camelus} in Fig. 
\ref{fig:histograms} is therefore difficult, and we leave this question for a follow-up study.
In any case, we seek to minimize the effect of the high-bias shaded bins on our cosmological 
analysis, and so we exclude them from the peak abundance summary statistic. This gives a data
vector for our ABC analysis that contains 93 total bins. This also has the effect of ensuring 
that the included bins all contain at least 10 peaks (horizontal dashed line), which is desirable 
for statistical purposes.

\subsection{Parameter inference with approximate Bayesian computation}
Approximate Bayesian computation (ABC) is an approach to constraining model parameters that avoids 
the evaluation of a likelihood function. With ABC in general, statistics of the observed data set 
are compared with the corresponding statistics derived from simulations assumed to model the process 
that generated the observations. For our purpose, the observed data set is the MICE catalogue,
and the simulations are generated by \textsc{Camelus}. Parameter space is then probed via 
accept-reject sampling, giving a fast and accurate estimate of the true parameter posterior 
distributions. 

As has been shown in \cite{LK.2015b} and \cite{LKP.2016}, ABC is an efficient and successful parameter
inference strategy for WL peak count analyses. The method has also been used recently in several 
other astrophysical and cosmological applications \citep{CP.2012, WSWV.2013, RRF.etal.2014, 
KBF.etal.2015, IVPL.etal.2015, ARA.etal.2015}. 
We constrain the parameter set $(\Omega_\mathrm{m}, \sigma_8, w_0^\mathrm{de})$ in this work, where $w_0^\mathrm{de}$ is 
the equation of state parameter of dark energy. $w_0^\mathrm{de}=-1$ corresponds to a pure cosmological 
constant. 

We implement a Population Monte Carlo (PMC) ABC algorithm to iteratively converge on the posterior 
distribution of parameters. The reader is referred to Algorithm 1 and the surrounding text in 
\cite{LK.2015b} for further details of PMC ABC as we implement it here. 
The algorithm proceeds by the following steps.
\begin{itemize}
  \item Draw an initial set of samples (called particles) from the prior distributions of the 
        parameters. We assume flat priors of $[0.1,0.9]$ for $\Omega_\mathrm{m}$, $[0.3,1.6]$ for 
        $\sigma_8$, and $[-1.8,0]$ for $w_0^\mathrm{de}$.
  \item For each particle, generate a data vector from one simulation of \textsc{Camelus} with its
        parameters set to those of the particle's location in parameter space.
  \item Compute the distance, defined as 
        \begin{equation}
          D(\bm{x},\bm{x}^\mathrm{obs}) = \sqrt{(\bm{x}-\bm{x}^\mathrm{obs})^T\bm{C}^{-1}(\bm{x}-\bm{x}^\mathrm{obs})},
          \label{eq:distance}
        \end{equation}
        between the proposed parameter set and the observed MICE data vectors. Here, $\bm{x}$ is
        the model prediction data vector, $\bm{x}^\mathrm{obs}$ is the MICE data vector, and 
        $\bm{C}$ is the covariance matrix. We note that due to the bias found from modeling
        (see Fig. \ref{fig:sig8_vs_Om_biased}), we also perform 
        constraints with the calibrated data vector. In this case, we add to the MICE data vector
        the difference between it and the model prediction computed with the MICE input parameters.
        See further details in Sect. \ref{sec:peak_constraints}.
  \item Discard the particles whose distance from MICE is larger than a prescribed tolerance 
        $\epsilon$.
  \item Reduce $\epsilon$ and iterate the process until the particle system representing the 
        posterior distribution converges.
\end{itemize}

Computing the distance requires the covariance matrix for the peak abundance data vector, which we 
estimate from 500 \textsc{Camelus} runs. We assume that $\bm{C}$ does not vary with cosmology and 
estimate it under $(\Omega_\mathrm{m},\sigma_8,w_0^\mathrm{de}) = (0.25,0.8,-1.0)$. We compute an 
unbiased inverse covariance estimator $\widehat{\bm{C}^{-1}}$ with a correction factor of 
$(n-p-2)/(n-1)=0.81$, where $n$ is the number of realizations, and $p$ the number of bins 
\citep{HSS.2007}. \cite{SH.2016} have shown that although this scaling indeed de-biases 
$\bm{C}^{-1}$, retaining a Gaussian likelihood is not correct. However, our large $n$ value 
ensures that $(\bm{x}-\bm{x}^\mathrm{obs})^T\widehat{\bm{C}^{-1}}(\bm{x}-\bm{x}^\mathrm{obs})$ 
is a good approximation to what should properly be a modified multivariate $t$-distribution.

\begin{figure}
\resizebox{\hsize}{!}{\includegraphics{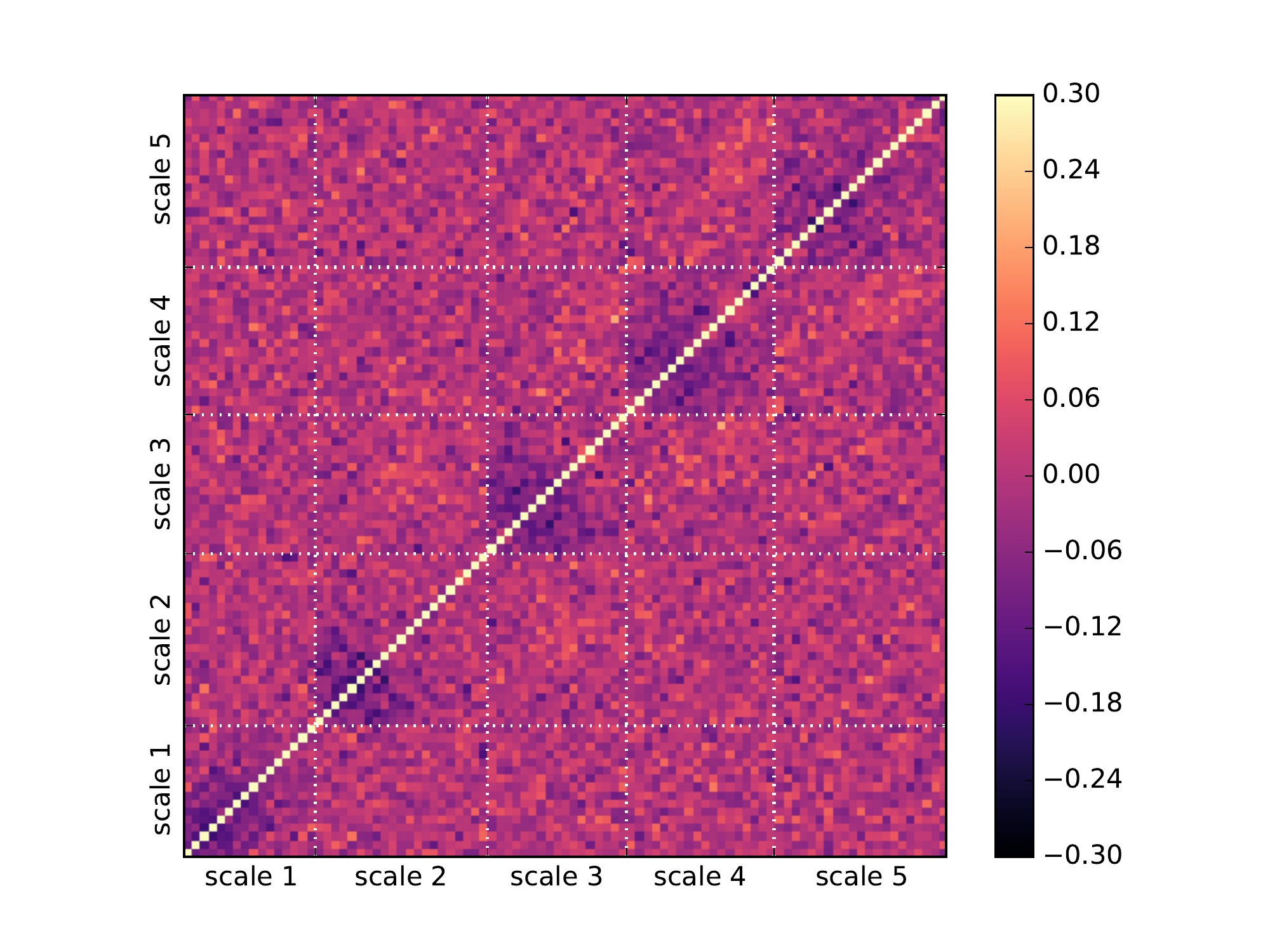}}
\caption{Correlation coefficients for our multiscale peak abundance data vector, computed from
         500 \textsc{Camelus} runs with $(\Omega_\mathrm{m}, \sigma_8, w_0^\mathrm{de}) = (0.25, 0.8, -1.0)$.
         Dotted lines delineate blocks corresponding to different wavelet scales. Correlation is
         weak both among bins within a particular scale, as well as across the different scales
         themselves. The latter reflects the efficient separation of multiscale information by 
         the starlet filter.}
\label{fig:camelus_corr}
\end{figure}

The associated correlation coefficients for our 93 S/N bins are shown in Fig. \ref{fig:camelus_corr}. 
Blocks delineated by dotted lines represent the correlation between the two corresponding wavelet 
scales. We see that S/N bins are not strongly correlated, and neither are different scales, 
reflecting the efficient multiscale separation of information by the starlet filter.

\section{Results}\label{sec:results}
We present in this section the cosmological parameter constraints attained from peak counts 
modeled by \textsc{Camelus} with ABC on the MICE field. We compare results with those of the 2PCF
statistics $\xi_\pm$.

\subsection{Parameter constraints from peak counts}\label{sec:peak_constraints}
\begin{figure}
\resizebox{\hsize}{!}{\includegraphics{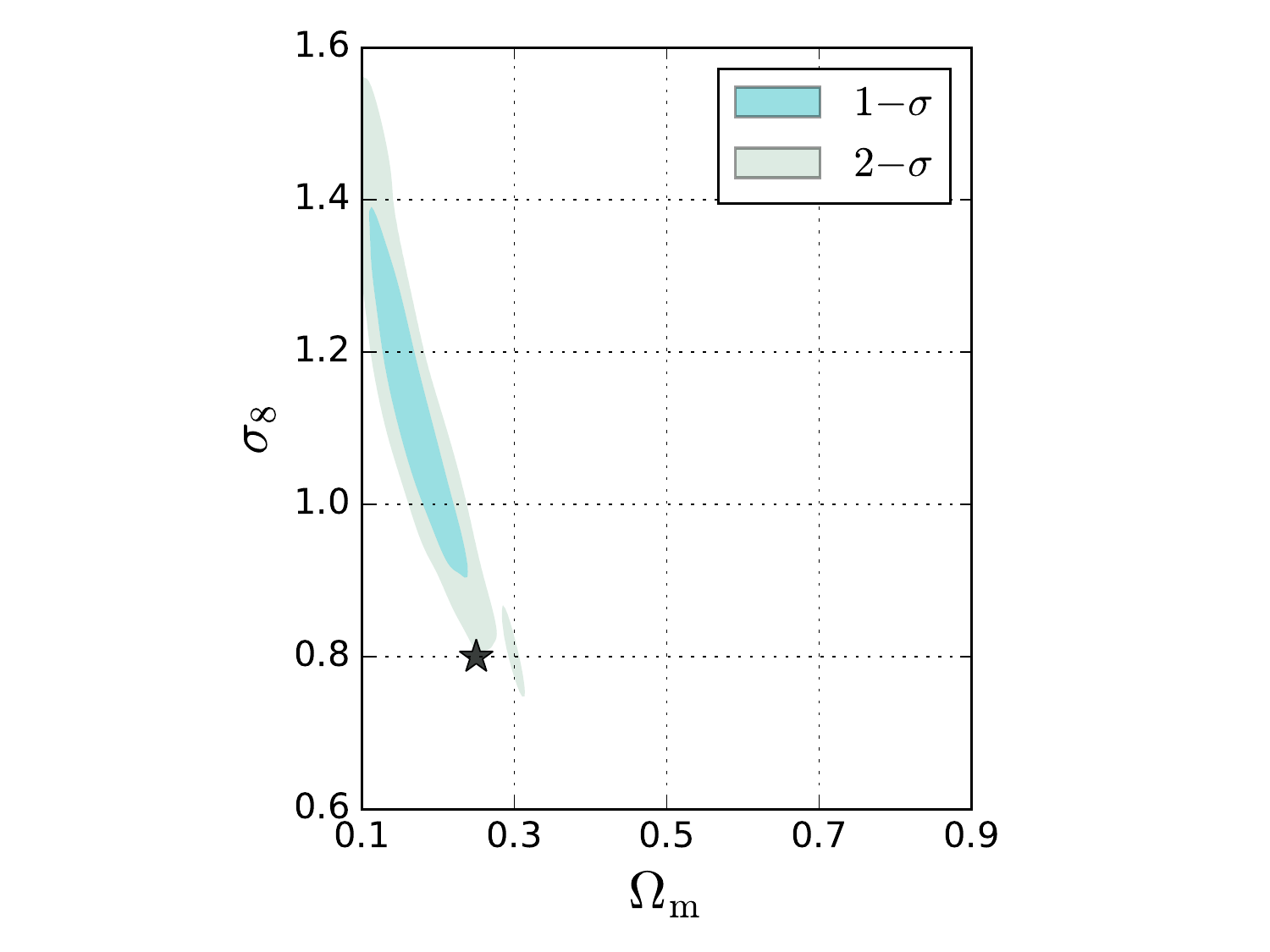}}
\caption{Parameter constraints in the $\sigma_8$--$\Omega_\mathrm{m}$ plane from peaks as modeled
         by \textsc{Camelus}. The 1-$\sigma$ and 2-$\sigma$ contours represent the 68.3\% and 
         95.4\% most probable regions, respectively, after marginalizing over $w_0^\mathrm{de}$.
         The full $\Omega_\mathrm{m}$ prior range is shown, and the star marks the MICE input
         cosmology. The constraint is biased significantly toward high $\sigma_8$ and low
         $\Omega_\mathrm{m}$ along the degeneracy direction.}
\label{fig:sig8_vs_Om_biased}
\end{figure}

\begin{figure*}
\centering
\includegraphics[width=17cm]{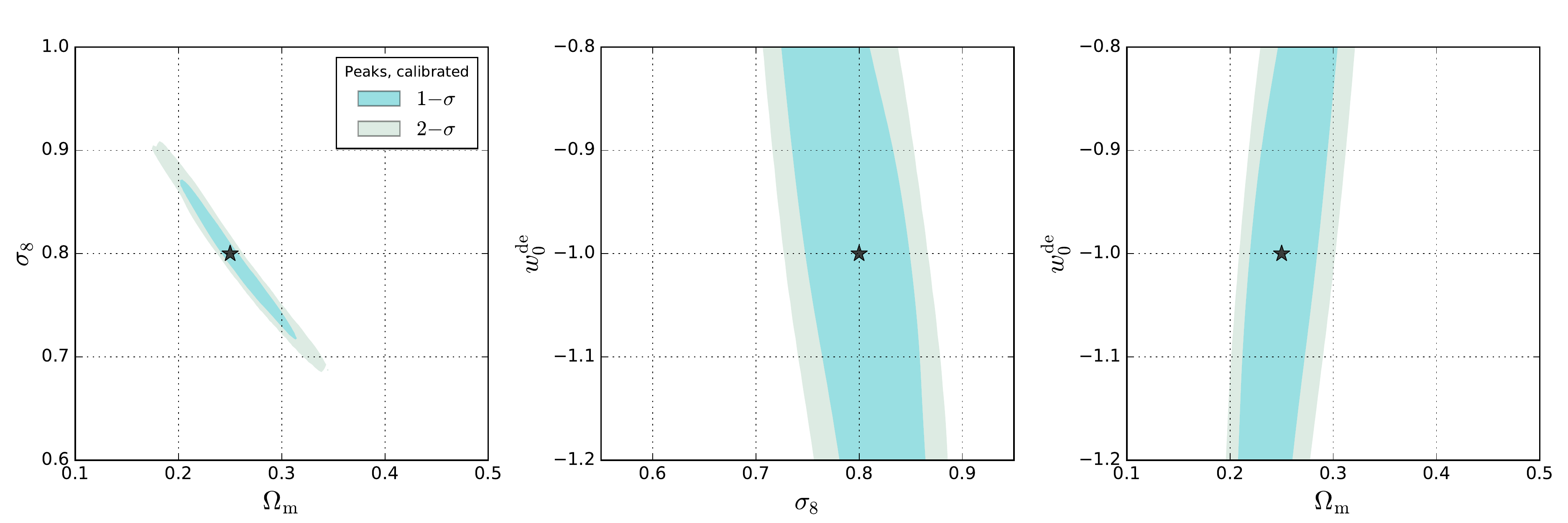}
\caption{Credible parameter contours from calibrated peak counts data vector for the three 2D 
         combinations of $(\Omega_\mathrm{m}, \sigma_8, w_0^\mathrm{de})$. 
         In each figure, the third parameter has been marginalized over, and the star 
         indicates the MICE cosmology. The left plot of $\sigma_8$--$\Omega_\mathrm{m}$ 
         space shows a right constraint with 1-$\sigma$ width measuring $\Delta\Sigma_8=0.05$. 
         The middle and right plots show that we cannot constrain $w_0^\mathrm{de}$ by our 
         analysis, but the degeneracy directions are still identifiable.}
\label{fig:contours_peaks}
\end{figure*}
The results of an ABC analysis are not point estimates of maximum likelihood, but rather credible
regions of parameter space. We show the 1-$\sigma$ and 2-$\sigma$ credible contours in the 
$\sigma_8$--$\Omega_\mathrm{m}$ plane after 12 iterations of ABC in Fig. \ref{fig:sig8_vs_Om_biased}. 
The areas within these contours represent the 68.3\% and 95.4\% most probable regions, 
respectively, after marginalizing over $w_0^\mathrm{de}$. With the PMC iteration scheme, ABC 
converges asymptotically to the true posterior. 
In practice, the contours stabilize quickly, meaning they do not continue to shrink
appreciably after only a few iterations. In this case, we stop after iteration 12, 
since the size and width of the contours are not significantly different compared to iteration 11.

Kernel density estimation (KDE) was used on the final ABC particle system to derive the contours. 
The expression for the multivariate KDE posterior is
\begin{equation}
  \hat{P}(\bm{x}) = \frac{1}{N}\sum_{k=1}^{N} W_{\bm{H}}(\bm{x}-\bm{x}^k),
\end{equation}
where $N=800$ is the number of particles, $\bm{x}^k$ is the $k$th sample, and $W_{\bm{H}}$ is the 
multivariate normal kernel with bandwidth $\bm{H}$ satisfying Silverman's rule 
\citep{Silverman.1986}.

It is well known that WL alone does not well constrain $\sigma_8$ and $\Omega_m$ separately due 
to the strong degeneracy between the parameters, and so should be combined with other cosmological 
probes with orthogonal contours. What we look at therefore is the thickness the contours in the 
direction of the degeneracy. This is frequently expressed by the derived quantity 
\begin{equation}
  \Sigma_8=\sigma_8(\Omega_\mathrm{m}/0.27)^\alpha,
\end{equation}
along with $\Delta\Sigma_8$, the 1-$\sigma$ uncertainty. 
The parameter $\alpha$ represents the best-fit slope in log space.
We find $\Sigma_8=0.77_{-0.05}^{+0.06}$ with $\alpha=0.75$, giving a 1-$\sigma$ width of 
$\Delta\Sigma_8=0.11$.

It is clear that despite having excluded the high-bias bins, the
residual \textsc{Camelus} systematics cause a significant bias in the parameter estimations. 
The peak contours are shifted along the degeneracy line so that the biases on $\Omega_\mathrm{m}$ 
and $\sigma_8$ are large, while that of $\Sigma_8$ is much reduced. Excluding the high-bias bins 
also weakens the constraining power, thereby broadening the contours. Whereas \cite{ZMHH.etal.2016} 
have found a good agreement between the \textsc{Camelus} algorithm and $N$-body runs on small 
fields, this test suggests that the systematics of \textsc{Camelus} need to be carefully studied 
in order to achieve accurate results for large-field analyses.

There are many possible sources of bias for peak predictions with \textsc{Camelus}. 
\cite{ZMHH.etal.2016} have already identified some concerning the fast halo modeling. These include
the lack of halo clustering, inaccurate modeling of the halo concentration, and the use of NFW
profiles. \cite{KKF.etal.2016} have argued as well that ignoring source clustering results
is a boost factor to peak counts. See also the discussion in Sect. \ref{sec:data_vector}
above.

As we are still interested in the constraining power of peaks with large-field statistics,
we explore further by supposing that the \textsc{Camelus} model can be improved and/or 
calibrated to correct the bias. One way to do this would be to first include effects like halo
clustering in the model, and then to tune the model parameters (e.g., of the mass function, NFW
profile, or mass-concentration relation) if necessary so that the peak abundance data vector 
produced by \textsc{Camelus} under the MICE cosmology matches the observation as closely as 
possible. A simpler variation, although less physically motivated, would be to just tune 
the model parameters without otherwise modifying the algorithm. Alternatively, one can calibrate 
directly at the level of the data vector, matching the peak count histograms of \textsc{Camelus} 
to MICE across all wavelet scales. We choose the latter approach here and leave a study of the 
former to a future publication.

In effect, calibrating this way is equivalent to pretending that the \textsc{Camelus} 
prediction under the MICE cosmology already matches the MICE observation. We note that 
in a more sophisticated treatment, and indeed in calibrating \textsc{Camelus} to use on real 
data, we would need to use many simulations with different cosmologies and interpolate by assuming, 
for example, that the calibration varies smoothly in parameter space.

We show the contours after calibration for the three combinations of 
$(\Omega_\mathrm{m},\sigma_8,w_0^\mathrm{de})$ in Fig. \ref{fig:contours_peaks}. The three panels 
represent results after 13 iterations of ABC, and the star again indicates the MICE input 
cosmology.
In the left plot are the credible contours in the $\sigma_8$--$\Omega_\mathrm{m}$ plane
(Note the different axis ranges compared to Fig. \ref{fig:sig8_vs_Om_biased}.)
We find now $\Sigma_8=0.76_{-0.03}^{+0.02}$ with $\alpha=0.65$. The 1-$\sigma$ width is therefore 
reduced to $\Delta\Sigma_8=0.05$, which agrees with the expectation that WL
is better able to distinguish $\sigma_8$ for larger $\Omega_\mathrm{m}$ values.
The posterior distribution of $\Sigma_8$ is shown as the blue curve in Fig. \ref{fig:Sigma8_PDF}. 
While the value of $\Sigma_8$ is not 
very meaningful in an absolute sense, it serves as a useful basis for comparing the tightness
of constraints attainable from different methods on the same data. We use 
$\Delta\Sigma_8$ to compare the amount of cosmological information contained in peaks to 
Gaussian probes of the WL signal in the following section.

The middle and right plots of Fig. \ref{fig:contours_peaks} show the $w_0^\mathrm{de}$ parameter 
spaces. The contours reveal that we cannot put constraints on $w_0^\mathrm{de}$ even with 
large-field surveys. This is consistent with the understanding that tomography is necessary to 
probe the late-time evolution of the Universe, for which $w_0^\mathrm{de}$ plays an important role. 
Nevertheless, the degeneracy between $w_0^\mathrm{de}$ and the other parameters is clear.

\subsection{Comparison with \texorpdfstring{$\xi_\pm$}{2PCF}}
We compute $\xi_\pm$ as defined by Eq. (\ref{eq:xi_pm}) on the same MICE sub-fields that were used 
for the peaks analysis with the publicly available 
Athena\footnote{\url{http://www.cosmostat.org/software/athena/}} 2D tree code.
We take the data vector in this case to be the concatenation of $\xi_+$ and $\xi_-$ computed in
10 log-spaced angular bins each. The covariance matrix is computed using the 186 extracted patches
from MICE, which we take to be independent realizations of 25 deg${}^2$ lensing fields. That is,
our estimator essentially measures the sub-sample covariances and then re-scales the result for 
the full survey \citep{NBG.etal.2009, FSE.etal.2016}.
\begin{figure}
\resizebox{\hsize}{!}{\includegraphics{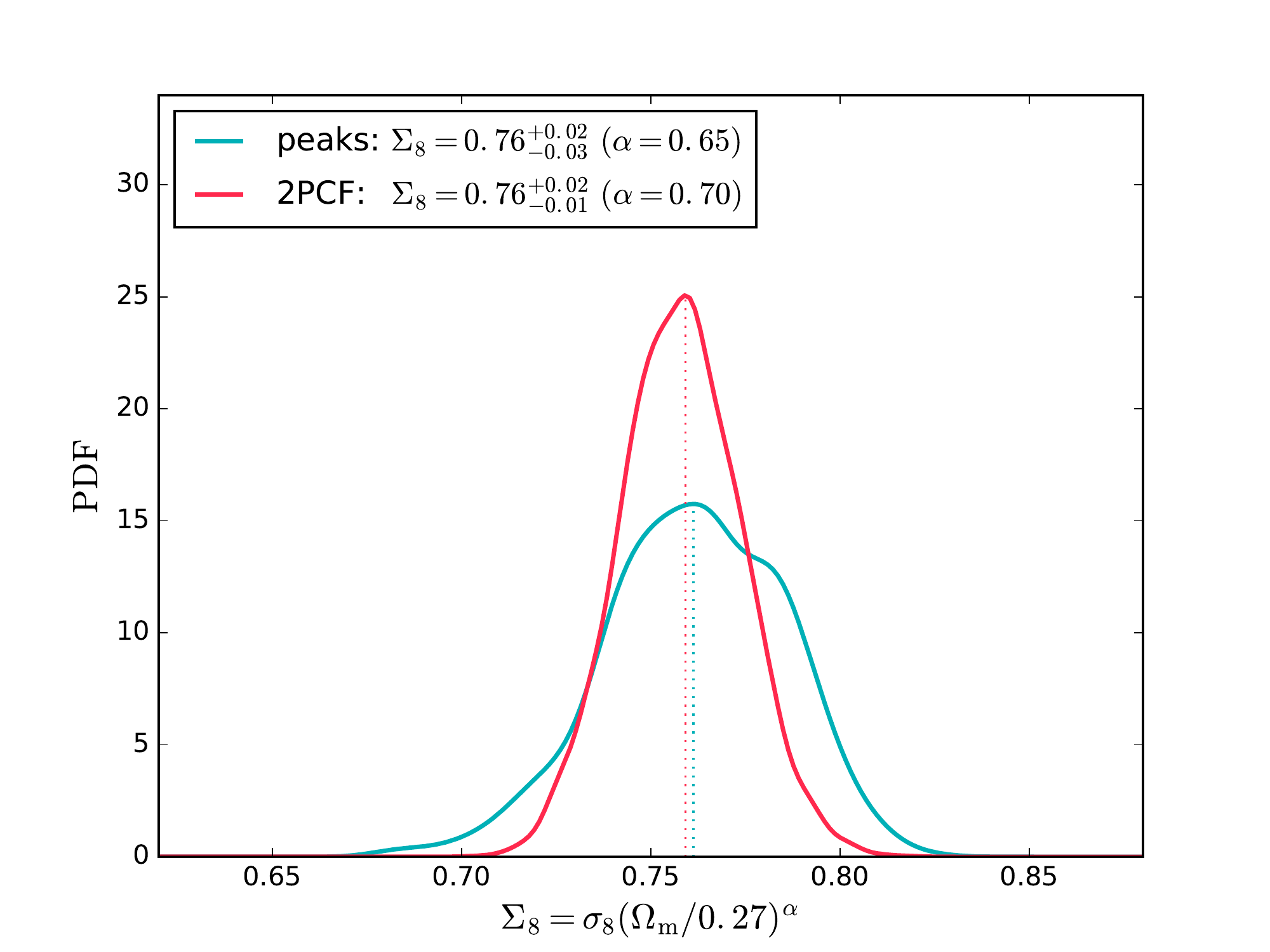}}
\caption{Probability density functions (PDFs) of the derived parameter $\Sigma_8$ from the 
         peak analysis and the two-point correlation functions $\xi_\pm$ (2PCF). For peaks, the 
         mode is $\Sigma_8=0.76$ with $\Delta\Sigma_8=0.05$, while for $\xi_\pm$, these are 0.76 and 0.03, 
         respectively. The comparable values of $\Delta\Sigma_8$ indicate the similar constraining
         power of the two WL statistics in a large-field survey.}
\label{fig:Sigma8_PDF}
\end{figure}

Estimating the covariance matrix in this way ignores correlations on scales comparable to 
the size of the patches, as well as smaller-scale separations spanning the boundaries between
patches. With the large number of patches and a maximum angular scale of 3 deg, we do not 
expect these caveats to significantly affect the results. It is likely, however, that we
underestimate the covariance to some degree due to the patches' not being perfectly independent.

\begin{figure}
\resizebox{\hsize}{!}{\includegraphics{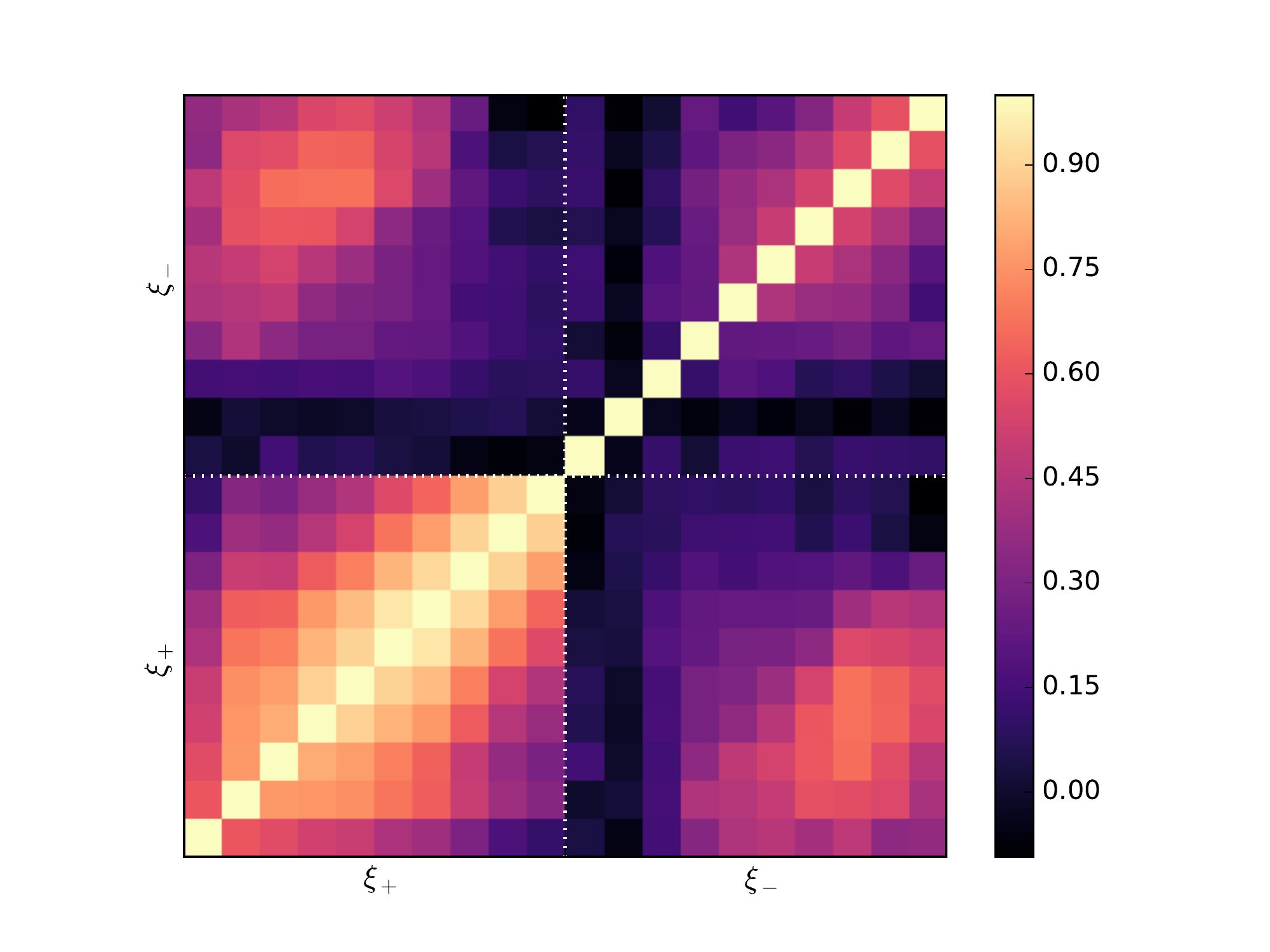}}
\caption{Correlation coefficients for $\xi_\pm$. Bins 1--10 correspond to $\xi_+$, and 11--20
         correspond to $\xi_-$. The covariance matrix of the full survey is computed as a
         re-scaling of the sub-sample covariance of the 186 MICE patches.}
\label{fig:xipm_corr}
\end{figure}

We show the correlation coefficients associated with $\xi_\pm$ in Fig. \ref{fig:xipm_corr}. Bins
1--10 correspond to $\xi_+$, and bins 11--20 correspond to $\xi_-$. The angular bins are evenly
spaced logarithmically between $(\theta_\mathrm{min}, \theta_\mathrm{max})=(1, 180)$ arcmin. There
is strong correlation among the $\xi_+$ bins and little to moderate correlation among the $\xi_-$
bins and their cross correlations with $\xi_+$.

As we do not have an analogous stochastic forward-model to predict $\xi_\pm$ as we do for peak 
counts, we cannot use ABC for parameter inference. Instead we use the population Monte Carlo 
software package CosmoPMC\footnote{\url{http://www2.iap.fr/users/kilbinge/CosmoPMC}}
\citep{WKB.etal.2009, KWR.etal.2010}. It is a Bayesian algorithm that uses adaptive-importance 
sampling to improve its estimation of the posterior distribution iteratively. For computing 
cosmology, the fitting formula used for the nonlinear matter power spectrum is the
halofit model of \cite{SPJ.etal.2003}. We use 
KDE in the same way as for the ABC result to analyze the resulting posterior particle system.

\begin{figure*}
\centering
\includegraphics[width=17cm]{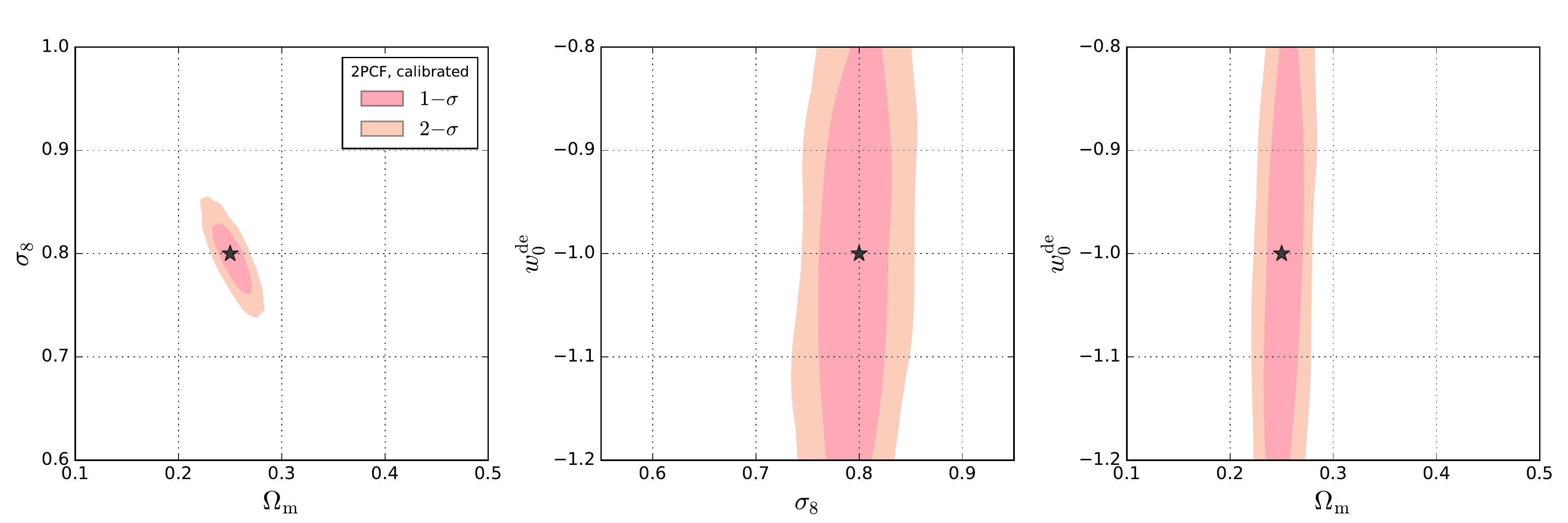}
\caption{Credible parameter contours from $\xi_\pm$ (calibrated) with CosmoPMC for the three 
         combinations of $(\Omega_\mathrm{m}, \sigma_8, w_0^\mathrm{de})$. 
         The width $\Delta\Sigma_8$ of the $\sigma_8$--$\Omega_\mathrm{m}$ contour is comparable 
         to, but in fact larger than the value from peaks, despite the distributions seen in Fig. 
         \ref{fig:Sigma8_PDF}. As with peaks, $w_0^\mathrm{de}$ cannot be constrained without a 
         tomographic analysis. The star indicates the MICE cosmology.}
\label{fig:contours_xipm}
\end{figure*}

The credible parameter contours for the calibrated 2PCF are shown in Fig. \ref{fig:contours_xipm}, 
analogous to Fig. \ref{fig:contours_peaks} for peaks. 
In the $\sigma_8$--$\Omega_\mathrm{m}$ plane, the 1- and 2-$\sigma$ contours are wider but less 
elongated compared to the peaks result. The degeneracy
direction is also slightly offset from that of peaks. The middle and right panels show that 
despite the increased statistical power of the large survey area, as with peaks, $w_0^\mathrm{de}$
cannot be constrained without tomography.

The $\Sigma_8$ posterior distribution for 2PCF is shown as the red curve in Fig. 
\ref{fig:Sigma8_PDF}, with $\Sigma_8=0.76_{-0.01}^{+0.02}$ and $\alpha=0.70$.
At first glance, it appears that 2PCF yields a tighter constraint on 
$\Sigma_8$. However, $\Delta\Sigma_8$ here does not reflect the real contour width from both left
panels of Figs. \ref{fig:contours_peaks} and \ref{fig:contours_xipm}. In fact, the contour width
from peaks is clearly smaller than from 2PCF. The discrepancy is due to the definition of 
$\Sigma_8$. Since it defines a power-law curve, it is not a good fit to the contour of Fig.
\ref{fig:contours_peaks}, which is elongated and barely bent. If one therefore looks to optimize
the synergy between WL and other probes to lift degeneracy, $\Sigma_8$ could be misleading. We
also point out that \cite{ZMHH.etal.2016} have found that the peak count covariance of 
\textsc{Camelus} is underestimated. This means that both the contours in the left panel of Fig.
\ref{fig:contours_peaks} and the distribution of $\Sigma_8$ in Fig. \ref{fig:Sigma8_PDF} should
be wider.

In the literature, similar comparisons have been done by \cite{DH.2010} using $N$-body simulations
and by \cite{LPH.etal.2015} using CFHTLenS data. In both studies, the authors found that peak-only
constraints are tighter than 2PCF-only constraints. The reason that our study does not show the
same tendency is due to the conservative cut of high-bias bins (see Fig. \ref{fig:histograms}),
which also removes cosmological information. A similar problem exists for the choice of scale
cutting for 2PCF. The real capacity to extract cosmological information depends strongly on the
accuracy of modeling, especially at nonlinear scales. Our study indicates that the tilt between
the respective degeneracy lines of the two observables can be significant. Peaks and 2PCF are 
therefore complementary, as their combination facilitates the breaking of parameter degeneracies.

\section{Summary and conclusions}\label{sec:conclusion}
Peak count statistics in weak lensing maps provide an important probe of the large-scale 
structure of the Universe. Peaks access the non-Gaussian information in the weak lensing signal, 
which is not captured by two-point statistics and the related power spectrum. The upcoming 
\textit{Euclid} mission will survey nearly one-third of the extragalactic sky and measure galaxy 
shapes with unprecedented precision, making it ideal for weak lensing analyses. 

To prepare for such next-generation data, we have studied the ability of peak counts to constrain 
the parameter set $(\Omega_\mathrm{m},\sigma_8,w_0^\mathrm{de})$ in a large-area simulation with 
\textit{Euclid}-like settings. We used the 5\,000 deg${}^2$ MICECATv2.0 lensing catalogue as mock 
observations, which we divided into 186 square 25 deg${}^2$ patches under gnomonic projection. 
Based on our cutting of the MICE octant, we extracted 4\,650 deg${}^2$ of effective area for the 
analysis. A more sophisticated scheme could be explored to make maximum use of the data, perhaps
by computing quantities directly on the sphere. The area was already large enough, however, to see
the effect of the increased statistical power on parameter constraints that we can expect from 
future surveys.

We used the stochastic forward-model code package \textsc{Camelus} to predict peak abundances 
as a function of cosmology. The model is semi-analytic, requiring only a halo profile and a 
halo mass function as input to generate a 25 deg${}^2$ lensing catalogue in a few seconds. We 
implemented a wavelet-based multiscale mass map filtering scheme in order to take advantage of 
cosmological information encoded at different scales. We showed that the 
\textsc{Camelus} model provides good agreement with the MICE peak histograms across five smoothing
scales for S/N values less than $\sim$5. On the other hand, for larger S/N values, we found an
increasing systematic deviation from MICE with increasing S/N.

Applying a conservative cut to exclude the highest bias S/N bins, we used approximate Bayesian 
computation to constrain the parameters of the MICE mock survey with peaks. In the 
$\sigma_8$--$\Omega_\mathrm{m}$ plane, we found tight 1- and 2-$\sigma$ contours oriented along 
the typical degeneracy direction seen in weak lensing analyses. We measured the derived parameter
$\Sigma_8=\sigma_8(\Omega_\mathrm{m}/0.27)^\alpha$ to be $0.77_{-0.05}^{+0.06}$ with best-fit 
power law slope $\alpha=0.75$ for a flat $\Lambda$CDM model. The uncertainty $\Delta\Sigma_8$
is representative of the width of the 1-$\sigma$ contour.

Despite omitting the high-bias S/N bins from the peaks data vector, the residual systematics in 
\textsc{Camelus} resulted in a large bias in the contours from the MICE input cosmology. The 
contours were shifted toward higher $\sigma_8$ and lower $\Omega_\mathrm{m}$ along the degeneracy 
direction.
This indicates the need to carefully study the systematics of \textsc{Camelus} before it can be
applied to real large-field data. We expect contributing systematics to include the lack of halo 
clustering, inaccurate modeling of the halo concentration, and the limitations of NFW profiles. 
It has been shown recently that these are a minor issue for relatively small fields 
\citep{ZMHH.etal.2016}, but the assumptions of \textsc{Camelus} clearly break down for large-area 
surveys.

To compare peak count results to those from Gaussian probes of WL, we computed the shear two-point 
correlation functions (2PCF) $\xi_\pm$ on the MICE field. Lacking a similar fast stochastic 
prediction algorithm for 2PCF as we have for peaks, we used the population Monte Carlo software 
CosmoPMC instead of ABC for parameter inference. For the comparison, we corrected for the 
\textsc{Camelus} bias by calibrating the peaks data vector. We also corrected for a smaller
bias found in the 2PCF result before comparing to peaks.

The comparison revealed comparable constraints in the $\sigma_8$--$\Omega_\mathrm{m}$ plane for both WL
observables, as measured by the width of their 1-$\sigma$ contours. We found $\Delta\Sigma_8=0.05$
for peaks and $\Delta\Sigma_8=0.03$ for 2PCF, although the actual
width of the contours was larger for 2PCF than for peaks. The reason for the apparent discrepancy
stems from the definition of $\Sigma_8$, as neither set of contours was particularly well fit by 
a power law. Nevertheless, the constraints of the two observables followed different degeneracy 
directions, indicating the benefit of a combined analysis with the two probes.
Neither peak counts nor 2PCF was able to constrain $w_0^\mathrm{de}$ without tomography.

We leave a tomographic study of peaks with \textsc{Camelus} to future work.
We have also not considered here certain realistic observational effects, such as intrinsic
alignment and masks. A primary benefit of the \textsc{Camelus} approach lies in its ability to
probe the true underlying PDF of observables. Its flexibility as a forward model will make it 
straightforward to include such effects after other systematics have been addressed. 

\begin{acknowledgements}
This work is supported in part by \textit{Enhanced Eurotalents}, a Marie Sk{\l}odowska-Curie 
Actions Programme co-funded by the European Commission and Commissariat {\`a} l'{\'e}nergie 
atomique et aux {\'e}nergies alternatives (CEA).

The authors acknowledge the Euclid Collaboration, the European Space Agency, and the support of 
the Centre National d’Etudes Spatiales (CNES).

This work is also funded by the DEDALE project, contract no. 665044, within the H2020 Framework 
Program of the European Commission.

The MICE simulations have been developed at the MareNostrum supercomputer (BSC-CNS) thanks to grants 
AECT-2006-2-0011 through AECT-2015-1-0013. Data products have been stored at the Port d'Informaci{\'o} 
Cient{\'i}fica (PIC), and distributed through the CosmoHub webportal (cosmohub.pic.es). Funding for 
this project was partially provided by the Spanish Ministerio de Ciencia e Innovacion (MICINN), 
projects 200850I176, AYA2009-13936, AYA2012-39620, AYA2013-44327, ESP2013-48274, ESP2014-58384, 
Consolider-Ingenio CSD2007- 00060, research project 2009-SGR-1398 from Generalitat de Catalunya, 
and the Ramon y Cajal MICINN program.

Finally, the authors would like to thank Peter Schneider, Sandrine Pires, and Samuel Farrens for 
useful comments and discussions.
\end{acknowledgements}

\bibliographystyle{aa}
\bibliography{refs}

\end{document}